\documentclass[12pt, openany, oneside]{memoir}

\usepackage[a4paper,left=2cm,right=2cm,top=2.5cm,bottom=4cm]{geometry}

\usepackage[utf8]{inputenx}
\usepackage[T1]{fontenc}

\usepackage{needspace}
\usepackage{marginnote}

\usepackage{microtype}
\usepackage{booktabs}

\usepackage{lmodern}           
\usepackage[scaled]{beramono}  
\usepackage{microtype}  

\makeatletter
\renewcommand{\maketitle}{\bgroup\setlength{\parindent}{0pt}
	\begin{flushleft}
		\large\bfseries{\@title}
	\end{flushleft}\egroup
}
\makeatother
\usepackage[explicit]{titlesec}
\usepackage{tikz}

\newcommand\titlebar{%
	\tikz[baseline,trim left=0cm,trim right=3cm] {
		\node [
		fill=yellow!90,
		text = black,
		anchor= base east,
		minimum height=3.5ex] (a) at (3cm,0) {
			\textbf{\thechapter}
		};   
	}%
}
\titleformat{\chapter}{\Large\bfseries\sffamily}{\titlebar}{0.25cm}{\textcolor{black}{#1}}      
\titlespacing*{\chapter}{-2cm}{3.5ex plus 1ex minus .2ex}{2.3ex plus .2ex}
\titlespacing*{name=\chapter,numberless}{-0cm}{3.5ex plus 1ex minus .2ex}{2.3ex plus .2ex}

\newcommand\sectionbar{%
	\tikz[baseline,trim left=0cm,trim right=3cm] {
		\node [
		fill=yellow!90,
		text = black,
		anchor= base east,
		minimum height=3.5ex] (a) at (3cm,0) {
			\textbf{\thesection}
		};   
	}%
}

\titleformat{\section}{\large\bfseries\sffamily}{\sectionbar}{0.25cm}{\textcolor{black}{#1}}      
\titlespacing*{\section}{-2cm}{3.5ex plus 1ex minus .2ex}{2.3ex plus .2ex}
\titleformat{\subsection}{\large\bfseries\sffamily}{\titlebar}{0.25cm}{\textcolor{black}{#1}}

\makeatletter
\renewcommand\@makefntext[1]{\leftskip=0em\hskip0em\@makefnmark#1}
\makeatother
\usepackage{url}
\usepackage{graphicx}
\usepackage{imakeidx}
\usepackage[hidelinks]{hyperref}
\makeindex[intoc]

\newenvironment{conf-abstract}[4][]{
	{\renewcommand\textsuperscript[1]{}
	\noindent
	\section[#2 \textit{(#3)}]{#2}
	}
	{\textit{#3}\par}
	\smallskip
	\noindent
	{#4\par}
	\bigskip
}{
\bigskip
}

\newenvironment{conf-talk}[4][]{
	{\renewcommand\textsuperscript[1]{}
		\noindent
		\section[#2 \textit{(#3)}]{#2}
	}
	{\textit{#3 (#4)}\par}
	\bigskip
}{
	\bigskip
}

\usepackage{etoolbox}
\newcommand{\indexauthors}[1]{%
  \forcsvlist{\index}{#1}
}

\usepackage{lipsum}

\makeoddfoot{plain}{}{}{\thepage} 
\makepagestyle{pter}
\makeevenhead{pter}{}{}{\itshape\leftmark}
\makeoddfoot{pter}{}{}{\thepage}
\makeevenfoot{pter}{\thepage}{}{}
\chapterstyle{komalike}

\begin{document}

\thispagestyle{empty}

\begin{titlingpage}

\vspace*{10em}

{
	\fontsize{24}{32}
	\bfseries 
	\boldmath
	\sffamily
	\noindent 
	Software Engineering for Intelligent and\\ Autonomous Systems (SEfIAS) \par
}

\vspace{4em}

{
	\fontsize{16}{19}
	\bfseries 
	\boldmath
	\sffamily 
	\noindent 
	Report from the GI Dagstuhl Seminar 18343\\August 19--24 2018, Schloss Dagstuhl\par
}

\vspace{8em}

{
	\normalsize
	\boldmath
	\sffamily 
	\noindent 
	Edited By:
	\par
}

{
	\fontsize{16}{19}
	\bfseries 
	\boldmath
	\sffamily 
	\noindent 
	Simos Gerasimou\\
	Thomas Vogel\\
	Ada Diaconescu
	\par
}
\end{titlingpage}
\frontmatter
\thispagestyle{empty}

\vspace*{13em}

\sffamily
\noindent 
\emph{Editors} \\[0.2cm]
\hspace*{-3mm}
\begin{tabular}{lll}
	Simos Gerasimou\\
	Department of Computer Science\\
	University of York, UK\\	
	\texttt{simos.gerasimou@york.ac.uk} \\
	\\	
	Thomas Vogel   \\
	Department of Computer Science \\ 
	Humboldt-Universität zu Berlin, DE\\ 
	\texttt{thomas.vogel@informatik.hu-berlin.de}\\
	\\
	Ada Diaconescu \\
	Computer Science and Networks department\\
	Télécom ParisTech, FR\\
	\texttt{ada.diaconescu@telecom-paristech.fr}\\
\end{tabular}

\bigskip
\bigskip
\bigskip
\bigskip

\noindent 
\emph{Published online with open access}\newline
\bigskip
\emph{Publication date}
(April, 2019)

\bigskip

\noindent 
\emph{License}\\[0.2cm]
\includegraphics[width=0.75\marginparwidth]{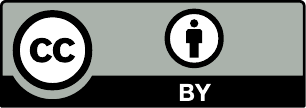}\newline
This work is licensed under a Creative Commons Attribution 3.0 Unported license (CC-BY~3.0): \texttt{http://creativecommons.org/licenses/by/3.0/legalcode}.\\
In brief, this license authorizes each and everybody to share (to copy, distribute and transmit) the work under the following conditions, without impairing or restricting the authors' moral rights:
\vspace{-1em}
\begin{itemize}
	\item Attribution: The work must be attributed to its authors.
\end{itemize}

\title{Report from the GI Dagstuhl Seminar 18343 \\
Software Engineering for Intelligent and Autonomous Systems}

\maketitle

\bigskip

\noindent
\textbf{Edited by}\\
Simos Gerasimou$^1$, Thomas Vogel$^2$, Ada Diaconescu$^3$
\ \\ 
\ \\$^1$ University of York, UK, simos.gerasimou@york.ac.uk\vspace*{1mm}
\ \\$^2$ Humboldt-Universität zu Berlin, DE, thomas.vogel@informatik.hu-berlin.de\vspace*{1mm}
\ \\$^3$ Télécom ParisTech, FR, ada.diaconescu@telecom-paristech.fr

\bigskip

\noindent
\textbf{Abstract}\\
\normalfont
Software systems are increasingly used in application domains characterised by uncertain environments, evolving requirements and unexpected failures; sudden system malfunctioning raises serious issues of security, safety, loss of comfort or revenue. 
During operation, these systems will likely need to deal with several unpredictable situations including variations in system performance, sudden changes in system workload and component failures. 
These 	situations can cause deviation from the desired system behaviour and require dynamic adaptation of the system behaviour, parameters or architecture. Through using closed-loop control, typically realized with software, \textit{intelligent and autonomous software systems} can dynamically adapt themselves, without any or with limited human involvement, by identifying abnormal situations, analysing alternative adaptation options, and finally, self-adapting to a suitable new configuration. The SEfIAS GI Dagstuhl seminar brought together early-career researchers and practitioners from the research communities of 
SEAMS\footnote{Int. Symposium on Software Engineering for Adaptive and Self-Managing Systems:\\ \url{http://self-adaptive.org/seams}}, 
ICAC\footnote{Int. Conference on Autonomic Computing: 	\url{http://icac2017.ece.ohio-state.edu/}}/ICCAC\footnote{Int. Conference on Cloud and Autonomic Computing: \url{http://autonomic-conference.org/iccac-2017/}}, 
SASO\footnote{Int. Conference on Self-Adaptive and Self-Organizing Systems: \url{http://www.saso-conference.org/}}, 
Self-Aware Computing\footnote{\url{https://se.informatik.uni-wuerzburg.de/research/self_aware_computing/community/}}
and 
AAMAS\footnote{Int. Conference on Autonomous Agents and Multiagent Systems: 
	\url{http://www.aamas-conference.org/}}, 
providing a forum for strengthening interaction and collaboration between these communities.

\bigskip

\noindent
\textbf{2012 ACM Subject Classification}: D.2.10 [Software Engineering] Design, D.2.11 [Software Engineering] Software Architectures


\newpage
\tableofcontents*

\mainmatter

\chapter{Executive Summary}

\textit{Simos Gerasimou, Thomas Vogel, Ada Diaconescu}
\indexauthors{Gerasimou!Simos, Vogel!Thomas, Diaconescu!Ada}

\bigskip
\noindent
Software systems are increasingly used in application domains characterised by 
uncertain environments, evolving requirements and unexpected failures; sudden 
system  malfunctioning raises serious issues of security, safety, loss of 
comfort or revenue.  
Such domains span all areas of our lives, from smart cities and electrical 
grids through defense, healthcare, finance, transportation and robotics to 
social networks, online commerce and distributed gaming.
During operation, these systems might need to deal with  
unpredictable situations including variations in system performance, sudden 
changes in system workload and component failures. These situations can cause 
deviation from the desired system behaviour and require dynamic adaptation of 
the system behaviour, parameters or architecture 
[1, 2].

Since the early 2000s, as the ubiquity and complexity of software systems had 
increased rapidly, there has been much interest in applying artificial 
intelligence and closed-loop control to the self-management of systems besides 
to emulating human intelligence, for instance, in playing chess. This trend led 
to research on engineering \textit{intelligent and autonomous} software systems 
capable of dynamically adapting themselves without any or with limited human 
involvement. Through using closed-loop control, typically realized with 
software, these systems can autonomously identify abnormal situations, analyse 
alternative adaptation options, and finally, self-adapt to a suitable new 
configuration. 

The opportunities for research and innovation in autonomous systems are 
remarkable. Horizon 2020 European research roadmaps highlight that advances in 
such systems will have a tremendous impact on society, business and global 
economy [3]. Autonomous systems will enhance our daily life by contributing to 
safer transportation, efficient manufacturing, secure systems and improved 
healthcare but they will also influence our social life; for instance, with 
respect to ethical issues on autonomous decision making. 

Over the past years, several research communities have devoted significant 
efforts to devise methodologies, algorithms and frameworks for engineering 
autonomous computing systems. 
Some noteworthy examples include the 
SEAMS, ICAC/ICCAC, SASO, Self-Aware Computing and AAMAS communities. 
Irrespective of the incarnation of each community, either focusing on resolving 
engineering problems in software applications (e.g., SEAMS) or systems in 
general (e.g., SASO, ICAC/ICCAC) or on investigating new learning, reasoning 
and prediction techniques for such systems (e.g., Self-Aware Computing, AAMAS), 
the main objective remains the same, that is, to make computing systems more
\textit{intelligent and autonomous}. 

Despite the mutual interests, these communities typically participate in 
disjoint research forums such as workshops, conferences and working groups. Hence, 
they rarely have the opportunity to meet in a common venue. This 
GI-Dagstuhl seminar brought together early-career researchers from these 
communities to discuss related research efforts, to evaluate the state of the 
art in each community, and to envision the future of intelligent and autonomous 
systems. 
Thus, the seminar strengthened interaction and collaboration between these communities, and encouraged the exploration of synergies for advancing the state of the art.
A common theme across these communities is \textit{software engineering}, since 
software typically realizes the system component that controls the 
\textit{intelligent} and \textit{autonomous} behavior of these systems. 

\subsection*{References}
[1] R. de Lemos, H. Giese, H. Muller, M. Shaw, J. Andersson, L. Baresi, B. 
Becker, N. Bencomo, Y. Brun, B. Cikic, R. Desmarais, S. Dustdar, G. Engels, K. 
Geihs, K. M. Goeschka, A. Gorla, V. Grassi, P. Inverardi, G. Karsai, J. Kramer, 
M. Litoiu, A. Lopes, J. Magee, S. Malek, S. Mankovskii, R. Mirandola, J. 
Mylopoulos, O. Nierstrasz, M. Pezze, C. Prehofer, W. Schafer, W. Schlichting, 
B. Schmerl, D. B. Smith, J. P. Sousa, G. Tamura, L. Tahvildari, N. M. Villegas, 
T. Vogel, D. Weyns, K. Wong, and J. Wuttke.  
A second Research Roadmap. Software Engineering for Self-Adaptive Systems II, 
volume 7475 of LNCS, pages 1–32. Springer, 2013.

\vspace*{2mm}\noindent
[2] S. Kounev, X. Zhu, J. O. Kephart, and M. Kwiatkowska. Model-driven 
Algorithms and Architectures for Self-Aware Computing Systems (Dagstuhl Seminar 
15041). Dagstuhl Reports, 5(1):164–196, 2015.

\vspace*{2mm}\noindent
[3] European Union. Horizon 2020 Work Programme 2018-2020. Future and Emerging Technologies. 2018. Available online at \url{http://ec.europa.eu/research/participants/data/ref/h2020/wp/2018-2020/main/h2020-wp1820-fet_en.pdf}
\chapter{Key Topics on Software Engineering for Intelligent and Autonomous Systems}

\begin{conf-abstract}[]
{Optimization}
{Erik Fredericks, Ilias Gerostathopoulos, Christian Krupitzer, Thomas Vogel}
{}
\indexauthors{Fredericks!Erik, Gerostathopoulos!Ilias, Krupitzer!Christian, Vogel!Thomas}

\noindent
The objective of the discussion group ``Optimization'' was to analyze and discuss the current state of the art in using multi-objective / multi-criteria optimization to improve the behavior of self-adaptive systems, mainly (but not solely) for the planning of adaptation. To work out the requirements of optimization in self-adaptive systems, we focused on four use cases all having variant requirements regarding the optimization aspect:
\begin{itemize}
	\item CrowdNav [1]: an approach to adaptive routing of vehicles, i.e., 
	optimization of CrowdNav using Bayesian optimization [2];
	\item Platooning [3], a technique for building convoys of self-driving vehicles;
	\item Sapienz [4] for adaptive, intelligent testing of software;
	\item Remote data mirroring (RDM) [5,6], i.e., optimization techniques to 
	harden the RDM configuration against uncertainty [7].
\end{itemize}

The discussion of the use cases identified several parameters that influence the choice of a suitable adaptation technique such as the problem domains / characteristics, the types of optimization strategies, and strategies for modeling systems and optimization points. Further, the absence of recommendations on means for identifying the most suitable optimization technique was identified.

To overcome this issue, the group decided to work on a taxonomy on optimization for self-adaptive systems. A bottom-up approach is favored here: based on a literature research to identify common techniques for optimization in self-adaptive systems, the idea is to analyze and compare the performance of several optimization techniques in different settings and to identify through that challenges, lessons learned, and an initial version of a taxonomy with selection criteria for choosing an optimization strategy for adaptive systems. A corresponding publication following this approach using the CrowdNav [1] app by integrating optimization loops in rule-based adaptation has been submitted to SEAMS as a full research paper, with a followup survey on incorporating optimization in self-adaptive systems planned by the group.  Moreover, another submission to an IEEE theme issue is also planned.

\subsection*{References}
[1] Sanny Schmid, Ilias Gerostathopoulos, Christian Prehofer, Tomas Bures. Self-Adaptation Based on Big Data Analytics: A Model Problem and Tool. In Proceedings of the 12th International Symposium on Software Engineering for Adaptive and Self-Managing Systems (SEAMS'17), 2017, pp. 102-108. 

\vspace*{2mm}\noindent
[2] Ilias Gerostathopoulos, Christian Prehofer, Tomas Bures. Adapting a System with Noisy Outputs with Statistical Guarantees. In Proceedings of the 13th International Symposium on Software Engineering for Adaptive and Self-Managing Systems (SEAMS'18), 2018, pp. 58-68.

\vspace*{2mm}\noindent
[3] Krupitzer, C ; Breitbach, M ; Saal, J ; Becker, C ; Segata, M ; Cigno, R 
Lo: RoCoSys: A framework for coordination of mobile IoT devices. In Proceedings 
of the 2017 IEEE International Conference on Pervasive Computing and 
Communications Workshops (PerCom Workshops), 2017, pp. 485–490.

\vspace*{2mm}\noindent
[4] Ke Mao, Mark Harman, Yue Jia. Sapienz: Multi-objective Automated Testing 
for Android Applications. In Proceedings of the 25th International Symposium on 
Software Testing and Analysis, 2018, pp. 94-105.  

\vspace*{2mm}\noindent
[5] M. Ji, A. Veitch, and J. Wilkes. Seneca: Remote mirroring done write. In 
USENIX 2003 Annual Technical Conference, pages 253-268, Berkeley, CA, USA, June 
2003. USENIX Association

\vspace*{2mm}\noindent
[6] K. Keeton, C. Santos, D. Beyer, J. Chase, and J. Wilkes.  Designing for 
disasters. In Proceedings of the 3rd USENIX Conference on File and Storage 
Technologies, pages 59-62, Berkeley, CA, USA, 2004. USENIX Association.

\vspace*{2mm}\noindent
[7] Fredericks, Erik M. ``Automatically hardening a self-adaptive system against 
uncertainty''. Proceedings of the 11th International Symposium on Software 
Engineering for Adaptive and Self-Managing Systems. ACM, 2016.

\end{conf-abstract}

\begin{conf-abstract}[]
	{Learning in Collective Autonomous Systems}
	{Mirko D'Angelo, Simos Gerasimou, Sona Ghahremani, Johannes Grohmann, 
		Ingrid Nunes, Evangelos Pournaras, Sven Tomforde}
	{}
	\indexauthors{D'Angelo!Mirko, Gerasimou!Simos, Ghahremani!Sona, Grohmann!Johannes, Nunes!Ingrid, Pournaras!Evangelos, Tomforde!Sven}
	
\noindent
Self-adaptive systems (SAS)~[1] refer to systems that can change their behaviour at runtime to keep meeting their design goals. They can be structured as a single software system, with a feedback loop that allows it to adapt itself, or be composed of multiple self-adaptive parts. In the latter case, these parts are called \emph{agents}~[2], which can interact leading to an emergent behaviour of the system as a whole. We refer to SAS composed of multiple agents as \emph{collective} SAS.

SAS and, consequently, agents are often situated in dynamic environments. Therefore, it is typically infeasible to predict at design time all possible scenarios that can occur at runtime to provide these systems with pre-specified adaptations. \emph{Learning} can be used to allow such systems and agents to observe the environment and outcomes of their actions in order to improve the performance of an individual agent or the system as a whole [3]. However, in many SAS learning has been explored in an application-specific way, because both researchers and practitioners have little guidance regarding best practices to solve recurrent problems in this context.

The ultimate goal of the ``Learning in Collective SAS'' group is to provide guidelines on how learning can be used in the context of (collective) SAS. In order to achieve this, first, there is a need to understand the commonalities among different SAS, so that the group is able to identify guidelines that are useful across different systems. Based on this goal, three research questions are derived, listed as follows.
	
\begin{itemize}
	\item \textbf{RQ1}: What are the characteristics of collective SAS that impact on how agents learn?
	\item \textbf{RQ2}: What, when and how can individual agents (part of a collective SAS) learn? 
	\item \textbf{RQ3}: What learning techniques are suitable for each learning task?
\end{itemize}

During the seminar, the group focused on answering RQ1 and managed to take steps towards the remaining questions. Inspired by the framework proposed by Luck et al.~[4], three characteristics, or \textbf{dimensions}, associated with SAS were identified as relevant to properly design a learning solution, leading to a framework to characterise SAS.

The first dimension is \textbf{selfishness}, which indicates whether an individual system (or agent) has the goal to maximise its own utility, i.e. it is selfish, or to maximise the utility of the system as a whole (or society), i.e. it is altruistic. The second dimension is \textbf{autonomy}, which reflects the amount of agents, part of the system, that are autonomous. Agents are autonomous when there is no external control over its behaviour. The third and last dimension is \textbf{observability} (this term is still being revised), which indicates what can be observed from the other agents that are part of the society. When there is minimal observability, agents can observe only the external behaviour of its neighbours, while when there is maximal observability, agents can also observe and obtain information regarding the internal behaviour of its neighbours.

These three dimensions lead to a continuous 3-D space in which collective SAS can be located. They are orthogonal and each combination of the different values associated with these dimensions leads to different design choices regarding what, when and how agents should learn. 

Based on these three dimensions and its possible minimal and maximal values, the group focused on analysing the impact of SAS with particular characteristics on the learning process. For example, in a system composed of selfish and autonomous agents and that has minimal observability, the adopted learning algorithms should be those that converge fast because agents want to maximise their own utility and no information is shared. In this example, six characteristics associated with this kind of system were identified with the corresponding implications over the learning process.

Driven by the preliminary results achieved during the break out sessions, our group performed a multifaceted analysis of collective SAS with learning abilities to answer research questions RQ1 -- RQ3. Based on this analysis, we introduce a 3D framework that illustrates the learning aspects of collective SAS considering the dimensions of autonomy, knowledge access, and behaviour, and facilitates the selection of learning techniques and models. The outcome of this work has been accepted for publication at the 14th IEEE/ACM International Symposium on Software Engineering for Adaptive and Self-Managing Systems (SEAMS’19) [5].

\subsection*{References}
[1] Mazeiar Salehie and Ladan Tahvildari, Self-adaptive Software: Landscape and Research Challenges, ACM Trans. Auton. Adapt. Syst. May 2009, pp. 14:1--14:42.

\vspace*{2mm}\noindent
[2] Nicholas R Jennings, An Agent-based Approach for Building Complex Software Systems, Commun. ACM, April 2001, pp. 35--41

\vspace*{2mm}\noindent
[3] Arthur Rodrigues, Ricardo D. Caldas, Gena\'{\i}na Nunes Rodrigues, Thomas Vogel and Patrizio Pelliccione. A Learning Approach to Enhance Assurances for Real-time Self-adaptive Systems, In Proceedings of the 13th International Conference on Software Engineering for Adaptive and Self-Managing Systems (SEAMS'18), 2018, pp. 206--216.

\vspace*{2mm}\noindent
[4] Michael Luck, Munroe Steve, Ashri Ronald and Fabiola López y López. Trust and norms for interaction. In 2004 IEEE International Conference on Systems, Man and Cybernetics, vol. 2, pp. 1944-1949. IEEE, 2004.

\vspace*{2mm}\noindent
[5] Mirco D' Angelo, Simos Gerasimou, Sona Ghahremani, Johannes Grohmann, Ingrid Nunes, Evangelos Pournaras, Sven Tomforde. On Learning in Collective Self-adaptive Systems: State of Practice and a 3D Framework. In 14th IEEE/ACM International Symposium on Software Engineering for Adaptive and Self-Managing Systems (SEAMS’19), in print.

\end{conf-abstract}

\begin{conf-abstract}[]
	{Control of Complexity}
	{Aimee Borda, Ada Diaconescu, Lukas Esterle, Alessandro V. Papadopoulos, Martin
		Pfannemüller, Danilo Pianini, Roberto Rodrigues Filho}
	{}
	\indexauthors{Borda!Aimee, Diaconescu!Ada, Esterle!Lukas, Papadopoulos!Alessandro V., 
		Pfannemüller!Martin, Pianini!Danilo, Filho!Roberto Rodrigues}
	
\noindent
The breakout-group ``Control of Complexity'' discussed some of the main challenges that an
autonomous system might face when deployed in real-world environments, and some of the
key control logic requirements to help overcome such challenges. We considered a wide
range of different application areas for autonomous systems, including collaborative robotics in cyber-physical systems and factory networks, autonomous vehicles and transportation networks, web and cloud servers, smart homes, smart offices, smart neighbourhoods and power grids.

While there is a wide range of challenges that autonomous systems might face in these
application scenarios, we identified the following three \textbf{challenges} as fundamentally important, and focused our discussions around them:
\begin{itemize}
	\item \textbf{Dynamic environment}: the system has to operate under constantly changing conditions. This includes external aspects as well as internal resources and potential collaborators.
	\item \textbf{Open systems}: the systems will change over time, as new components are added while	others become obsolete, or fail.
	\item \textbf{Changing goals}: the system’s goals may change over time, either because they are redefined by stakeholders, or because they need to be readjusted to unforeseen
	situations.
\end{itemize}

To address the above challenges, the control logic of autonomous systems must meet the
following specific \textbf{requirements}. First, the control logic needs to be adaptable. The controller has to be able to adapt itself to changing conditions (i.e. external, internal and goal-related) to ensure the system operates as expected (i.e. meeting its latest stakeholder goals). Second, the controller needs to be scalable. This control logic requirement encompasses three aspects: (i) support for significant increases in the numbers of components and systems to be controlled, during runtime; (ii) ability to control systems and components with heterogeneous capabilities, resources, and behaviours; and (iii) ability to develop more complex solutions by combining simpler control logic elements, while managing potential conflicts. Third, the controller of an open system, operating in dynamic environments with changing goals, needs to be able to trade-off the ability to adapt and self-improve against the guarantees it can provide. In safety-critical situations, controllers should guarantee satisfactory system behaviour, even if its behaviour is sub-optimal. However, in non-critical situations, the controller should be able to explore new control strategies, aiming to optimise system performance, even if this might temporarily breach performance guarantees. 

To help meet these requirements we propose that software engineering methods and
artefacts should ensure the following set of generic control logic \textbf{capabilities}:
\begin{itemize}
	\item \textbf{Controller Composability}: 
	different control modules, encapsulating specific internal structure and behaviour, can be integrated at runtime -- e.g., cooperate to achieve overall system control and meet system goals. This requires that each controller provides an external representation describing its properties, goals, models, behavioural states, and interfaces. Such formal representation can be parsed and processed by other controllers and potentially by humans. Controller components need to be able to query and discover other controller components, as well as to test and validate their descriptions. The system should support controller composition either offline or online; and either by human operators or via autonomous facilities. Hence, system goals specification and transformation are highly relevant to our work -- this insight led to fruitful exchanges with another break-out group discussing systems able to autonomously manage own goals.
	
	\item \textbf{Control Architecture}: 
	Multi-scale (hierarchical) controller organisations manage different system scales. For instance, lower level controllers can perform urgent actions to react to environmental changes, while higher levels can perform more intricate planning, consider global system optimisation, and coordinate lower levels accordingly. To ensure system coherence, higher controllers should operate at lower frequencies than lower controllers. Such architecture facilitates frequent changes in a system’s controller composition, as it limits the impact of local changes on the overall system.
	
	\item \textbf{Adaptability of controllers}: 
	Control components can be hot-swapped, e.g., based on predefined, dynamically discovered control components. Furthermore, controller parameters and description models can be learned during runtime. This allows a controller to change its internal strategies and tune its achievable goals during runtime; and to better integrate with other controllers. An important objective for adaptive controllers is to maintain guarantees requested by stakeholders (at design-time or during runtime). Predefined and pre-tested component hot-swapping offers more guarantees but less adaptability to unforeseen conditions; while an intelligent, more ``creative'' controller may better deal with uncertain environments but ensure fewer guarantees.
\end{itemize}

Future research directions should provide software engineering artefacts to support the
above controller requirements and capabilities -- e.g. reusable component models,
architectures, frameworks, and design patterns; development methodologies; verification and
validation techniques. We additionally identified two specific research questions. First, can we offer a system design which guarantees that suitable local behaviour (e.g. achieving local goals, being optimal, meeting guarantees) will eventually lead to suitable system-wide
behaviour (e.g. achieving global goals, being optimal overall, meeting system guarantees)?
This would require either that controllers can be composed linearly so as to offer predictable composed control functions; or, that runtime side-effects emerging from dynamic controller integration can be detected and resolved during runtime, in a predictable way. Second, we wonder whether or not stateless control components would integrate more efficiently. This would require the system to deal with states progressively, throughout the hierarchical levels. Furthermore, this requires controllers to share and distribute knowledge among themselves, across hierarchical levels. 

Finally, we identified relevant concepts and approaches from related research fields --
including systems theory, control theory, self-* systems, and complex systems -- which may
prove valuable for tackling the identified challenges of autonomous and intelligent control of large-scale highly-adaptable systems. In future research we will further investigate these complementary techniques and mechanisms.

\end{conf-abstract}

\begin{conf-abstract}[]
	{Specification and Composition of Non-Functional Requirements with Capabilities for Self-Adaptive Systems}
	{Nico Hochgeschwender and Sebastian Götz}
	{}
	\indexauthors{Hochgeschwender!Nico, Götz!Sebastian}

\noindent
A plethora of approaches to develop and operate self-adaptive software systems has been introduced within the last two decades. In [1], Salehi and Tahvildari differentiate such systems into self-configuring, self-protecting, self-healing and self-optimizing systems. Here, self-configuring systems denote the most basic form of self-adaptiveness, by stating that the systems are able to adjust their own configuration. The three other classes further specify the purpose for which the system reconfigures itself: to heal, to protect or to optimize itself, usually w.r.t. specified non-functional requirements (NFR) such as performance or energy consumption. Interestingly, current approaches for self-optimizing systems use languages with limited expressiveness to specify NFRs. 

Typically, they are specified with fixed reference values. For example, a specific maximum latency or a concrete minimum bandwidth is specified. However, for the operation of real world systems like autonomous robots, drones or cars, such hard constraints are often not required and potentially narrow the solution space unnecessarily. In addition, those systems are expected to remain operational even in the presence of situations and conditions where minimum and maximum constraints are violated by, for example, varying task, platform or environment requirements. Therefore, the minimum or maximum constraints could be relaxed - a feature usually not supported by the language used to specify NFRs. By relaxing NFRs, the set of valid configurations increases, which could be leveraged by the constraint solver to find a valid solution more efficiently. In the literature, some notable approaches to express NFRs in a more relaxed way are available, for example, the work of Whittle et al. [2] proposes the RELAX language which provides a vocabulary to enable developers to state NFRs in an uncertain manner by explicitly handling uncertainty of requirements with temporal, ordinal and modal operators. However, the main objective of RELAX and related approaches is to specify what should be achieved in terms of requirements whereas means to describe the how, for example, system abilities which are contributing to achieve NFRs, is typically not available in those approaches. 

In our breakout group we investigated whether a unified modeling approach of both non-functional requirements and system abilities could enhance the development of autonomous systems capable of carrying out their tasks even in the presence of disturbances and failures. To this end, we propose to employ the Conceptual Space formalism introduced by Gärdenfors [3] and rooted in the cognitive science domain, to model both NFRs and system abilities. The general idea of this formalism is to bridge the gap between symbolic and sub-symbolic knowledge by introducing an intermediate, geometric representation layer, namely the Conceptual Space. In essence, a Conceptual Space is a high-dimensional space composed by a number of dimensions. Each dimension refers to a measurable unit (e.g., bandwidth, latency or throughput) and the convex regions formed in that space correspond to high-level concepts (e.g., requirements or system abilities). The geometric nature of the Conceptual Space formalism enables not only to construct arbitrary variations of concepts, but also to perform different operations supported by the underlying algebra [4]. First, the geometric nature allows to apply distance measures (e.g., euclidean distance) to compare different instances (points in the Conceptual Space) with each other. This feature is helpful in selecting system configurations which closely match specific NFRs [5] by assessing their distance value obtained from a distance measure. Second, recent developments of the algebra pave the way to apply operations to compute whether a certain NFR concept is provided by a system ability and to which degree. Further, the algebra provides means to combine different concepts with each other - a feature which could be exploited in scenarios where self-adaptive systems are required to compose their abilities in order to achieve common goals under specific NFRs.

\subsection*{References}
[1] Mazeiar Salehie and Ladan Tahvildari, Self-adaptive Software: Landscape and Research Challenges, ACM Trans. Auton. Adapt. Syst. May 2009, pp. 14:1--14:42.

\vspace*{2mm}\noindent
[2] Jon Whittle, Pete Sawyer, Nelly Bencomo, Betty Cheng and Jean-Michel Bruel. ``RELAX: Incorporating Uncertainty into the Specification of Self-Adaptive Systems''. In Proceedings of the 17th IEEE International Requirements Engineering Conference. 2009. pp. 79-88

\vspace*{2mm}\noindent
[3] Peter Gärdenfors. ``Conceptual spaces: The geometry of thought''. The MIT Press. 2004.

\vspace*{2mm}\noindent
[4] Lucas Bechberger and Kai-Uwe Kühnberger.  ``A Thorough Formalization of Conceptual Spaces''. In Proceedings of the Joint German/Austrian Conference on Artificial Intelligence. 2017. pp. 58-71

\vspace*{2mm}\noindent
[5] Sebastian Blumenthal, Nico Hochgeschwender, Erwin Prassler, Holger Voos and Herman Bruyninckx. ``An Approach for a Distributed World Model with QoS-based Algorithm Adaptation''. In Proceedings of the IEEE/RSJ International Conference on Intelligent Robots and Systems (IROS). 2015. pp. 1806-1811

\end{conf-abstract}

\begin{conf-abstract}[]
	{Goals}
	{Christian Cabrera, Sylvain Frey, Fatemeh Golpayegani, Barry Porter, Romina Spalazzese}
	{}
	\indexauthors{Cabrera!Christian, Frey!Sylvain, Golpayegani!Fatemeh, Porter!Barry,  Spalazzese!Romina}

\noindent
During the breakout sessions in SEfIAS Dagstuhl Seminar, we discussed the importance of
goal definition, goal translation and the connectivity of the goals of different roles defined within the system such as users, providers, designers etc. In this report, we are summarising the importance of this topic and also discussing a number of examples to exemplify this issue in real-world applications.

When designing self-adaptive systems in different context such as IoT, multi-agent systems,
or control systems, a crucial question that one should ask prior to design the system is ``what is the goal?'' The definition of the system goal and the way how it is formalised plays an important role during the life cycle of a self-adaptive system, from feasibility study to maintenance and the way how it will adapt and evolve. This is particularly important when we design systems that require to interact with humans, where should understand the need of a human, be able to translate these needs to an understandable machine language and also be able to respond to new or evolving needs when it has the required capabilities. This whole process requires the machine to understand the goals/needs at an abstract level, decompose it to a lower level set of sub-goals to make it possible for the computing entities (e.g., agents) to realize it. It should also have the ability to understand new goals and adapt its decomposing process to realize new/evolved goals.

The following challenges were identified during the discussions:
\begin{itemize}
	\item Translation of humans’ goal (end-users) to machine understandable goals
	
End-user goals are commonly expressed in natural or high-level languages that machines
cannot understand. Current methods of translation from human goals to machine
understandable goals implies high human intervention and cause loss of information that
systems cannot use in their internal processes. This information is usually valuable and
would allow systems to have better results and performance. Novel approaches to translate
human goals to machine understandable languages are needed. As a first step to
understand this problem we plan to:
	\begin{itemize}
		\item Identify different levels of abstraction from a goal defined by a human, to a
		realizable goal by a machine in different domains
		
		\item Identify/define the links between these levels, how the information is
	transformed from one level to another and which information is lost in these
	transformations
	
		\item Define methods to keep the lost information in the transformation and express
	it in an useful format for the system
	\end{itemize}

\item Goal adaptation at runtime with emergent capabilities

Systems have pre-defined goals and capabilities according to what is available in the
environment before execution. New capabilities can emerge or current ones disappear in
dynamic environments. Systems must adapt their goals and execution plans according to
the continuously changing set of available capabilities.

\end{itemize}

\end{conf-abstract}
\chapter{Overview of Talks}

\begin{conf-talk}[]
	{Compositional Modelling and Verification of Self-Adaptive Systems}
	{Aimee Borda}
	{Trinity College Dublin, IR}
	\indexauthors{Borda!Aimee}
\noindent
We are looking into a systematic and feasible approach for designing and verifying Self-Adaptive systems. We investigate different compositional techniques and models that can be verified using existing technologies.	
\end{conf-talk}

\begin{conf-talk}[]
	{Urban-Centric Service Discovery: A Self-Adaptable Model for Smart Cities}
	{Christian Cabrera}
	{Trinity College Dublin, IR}
	\indexauthors{Cabrera!Christian}
	\noindent
A smart city is an environment that is continuously changing where citizens cannot be constantly aware of all relevant services around them. The ICT systems that support a city should evolve with it to offer the right services to the right citizen, in the right place, at the right time. Some existing research proposes static and reactive solutions where service organisation and discovery is defined by network properties, the service domain, service usage, or the city context to offer efficient service discovery. However, they do not evolve according to the city dynamics. This lack of flexibility produces outdated distribution of services information that negatively impacts discovery efficiency as the city changes. Self-adaptable service discovery solutions have been also proposed but they do not scale well in large scenarios such as a smart city. We propose a self-adaptable service model for smart cities to support efficient and pervasive discovery based on urban-context and citizens' behaviour. This model reorganises services information according to city events, and offers both reactive and proactive service discovery depending on the environment status.
\end{conf-talk}

\begin{conf-talk}[]
	{Decentralized Self-Adaptive Computing at the Edge}
	{Mirko D'Angelo}
	{Linnaeus University, SE}
	\indexauthors{D'Angelo!Mirko}
	\noindent
Nowadays, computing infrastructures are usually deployed in fully controlled environments and managed in a centralized fashion. Leveraging on centralized infrastructures could prevent the system to deal with scalability and performance issues, which are inherent to modern large-scale data-intensive applications. We envision fully decentralized computing infrastructures deployed at the edge of the network providing the required support for operating data-intensive systems (DiS). However, engineering such systems raises many challenges, as decentralization introduces uncertainty, which in turn may harm the dependability of the system. The research directions and current contributions towards this vision address the following questions: (i) when is decentralized computing required in DiS, (ii) how to enable decentralized computing in DiS?, (iii) how to design/analyze decentralized DiS?.	
\end{conf-talk}

\begin{conf-talk}[]
	{Generic Architectures for Multi-Level Goal-driven Self-Integrating Systems}
	{Ada Diaconescu}
	{Télécom ParisTech, FR}
	\indexauthors{Diaconescu!Ada}
\noindent
Self-integration enables socio-technical systems to adapt to a wide range of internal and external changes, so as to achieve their goals (which may also change). Multi-level, or hierarchical designs, help system scalability (with respect to the number of components, their heterogeneity and frequency of change. The aim of my research is to provide reusable conceptual frameworks, architectures, design patterns and methodologies to help designers understand and develop large-scale highly-adaptive (socio-)technical systems, relying on self-integration and hierarchical designs.	
\end{conf-talk}

\begin{conf-talk}[]
	{Autonomous Decision Making for Collaborating Agents}
	{Lukas Esterle}
	{Aston University, UK}
	\indexauthors{Esterle!Lukas}
	\noindent
When autonomous agents interact, they have to make decision on an ongoing basis. These decisions can be based on information sensed from the environment or received from other agents. When they collaborate with each other towards common goals, these decisions need to be sensible in order to not affect these goals too much. My research interest is on one hand on the amount of information required to allow agents to arrive at reasonable decisions. This becomes more pressing when the environment can change and/or agents can move about in the world. On the other hand, I am interested in enabling agents to reflect on their decisions especially with respect on the impact on their own performance and on the performance of others.	
\end{conf-talk}

\begin{conf-talk}[]
	{Software Engineering for Self-Adaptive Cyber-Physical Systems}
	{Erik Fredericks}
	{Oakland University, USA}
	\indexauthors{Fredericks!Erik}
	\noindent
Cyber-physical systems have become ubiquitous, especially in such domains that are safety-critical in nature.  My research focuses on the implications of combining cyber-physical systems, self-adaptive systems, and search-based software engineering.  Specifically, I am interested in how uncertainty can impact such systems and how software engineering techniques can be used to enhance assurance at all stages of the software life cycle for those systems that are safety-critical.  I will discuss recent work in search heuristics for non-functional software requirements, a self-adaptive medical smart home, and run-time software validation.	
\end{conf-talk}

\begin{conf-talk}[]
	{Launch photon torpedos: a journey through organic cyberz}
	{Sylvain Frey}
	{DeepMind, UK}
	\indexauthors{Frey!Sylvain}
	\noindent
My research deals with complex, dynamic cyber-physical systems with critical safety and security requirements such as the Internet, smart grids, Industrial Control Systems, and critical infrastructures in general. My expertise spans various domains: cyber security, software engineering, multi-agent systems,	
\end{conf-talk}

\begin{conf-talk}[]
	{Assurances for AI-Based Systems}
	{Simos Gerasimou}
	{University of York, UK}
	\indexauthors{Gerasimou!Simos}
	\noindent
Autonomous systems can sense, reason, and interact with the real world. Recent advances in Artificial Intelligence and related technologies have raised our expectations for engineering fully-autonomous systems including driverless cars and interactive companions. While the expected societal, economic and safety benefits are significant, some very recent unfortunate incidents (e.g., Uber, Tesla) indicate that we are not there yet. In this talk, we will explore the opportunities coming with autonomous systems, investigate the challenges for engineering trustworthy autonomous systems and analyse what all this means to our everyday lives. 	
\end{conf-talk}

\begin{conf-talk}[]
	{Online Experiment-Driven Adaptation}
	{Ilias Gerostathopoulos}
	{Technical University Munich, DE}
	\indexauthors{Gerostathopoulos!Ilias}
	\noindent
As modern systems become larger, more complex and customizable, it is difficult to fully model their internal workings in advance in order to control and optimize them at runtime. Thus, we propose here optimization based on operational data and experimentation. In this talk, I will present the main ideas behind Online Experiment-Driven Adaptation (OEDA), an approach whereby systems are adapted and configurations are evaluated by controlled experiments in production environments using advanced data analysis and statistical methods. In particular, I will describe the main concepts of OEDA, along with an example of a complex, hard-to-model system that is amenable to online optimization via controlled experiments—the CrowdNav self-adaptation model problem. I will also describe three different types of costs in online experimentation and discuss the tradeoffs in terms of the different costs of using three different optimization algorithms: Bayesian optimization, factorial design, and local search.	
\end{conf-talk}

\begin{conf-talk}[]
	{Utility-driven self-adaptation of large dynamic architectures}
	{Sona Ghahremani}
	{Hasso Plattner Institute, DE}
	\indexauthors{Ghahremani!Sona}
	\noindent
Self-adaptation can be realized in many ways. In my research I propose a rule-based and utility-driven approach that achieves the beneficial properties of each of these directions such that the adaptation decisions are optimal while the computation remains scalable. The approach can be used for the architecture-based self-healing and self-optimization of large software systems. We define the utility for large dynamic architectures of such systems based on patterns capturing the issues self-adaptation must address. Therefore, self-adaptation can be steered by predicting changes of the system utility. However, construction of an analytic representation of the system utility is challenging due to lack of detailed information about the system performance model. We mitigate this problem with a methodology to learn the changes of the system utility without relying on detailed information of the system.	
\end{conf-talk}

\begin{conf-talk}[]
	{Collaboration Community Formation in Open Systems for Agents with Multiple Goals}
	{Fatemeh Golpayegani}
	{Trinity College Dublin, IR}
	\indexauthors{Golpayegani!Fatemeh}
	\noindent
Agents frequently coordinate their behaviour and collaborate to achieve a shared goal, share constrained resources, or accomplish a complex task that they cannot do alone. Forming an effective collaboration community in which agents are willing to cooperate, and have no conflict of interests, is the key to any successful collaborative process. Forming such communities has been addressed well in cooperative and closed multi-agent systems. However, it is particularly challenging in open multi-agent systems where agents are self-interested. Such agents are also likely to continuously and unpredictably leave and join the system and have multiple goals to pursue simultaneously.  Existing research has addressed this challenge in open systems with utility-based or complementary-based approaches. Utility-based approaches focus on maximising self-interested agents' individual pay-off when sharing constrained resources. In complementary-based approaches, agents' individual skills are composed to accomplish a complex task or achieve a shared goal. However, in such systems agents need to identify the possible dependencies and conflicts between their individual goals, and build/adapt collaboration communities to pursue multiple goals simultaneously. Such dependencies affect agents' levels of self-interest and consequently their willingness to form collaboration communities. Given the circumstances, agents need a decentralised mechanism to acquire an understanding of other agents operating in their system, identify their goal dependencies, and adapt their level of self-interest to form effective collaboration communities.  My research focus is on proposing and developing a fully decentralised approach to Collaboration Community FOrmation Model for agents with multiple goals in open systems (CCFOM). CCFOM presents a new social reasoning model and a new distributed community formation algorithm. CCFOM enables agents to pursue their individual and shared goals simultaneously in resource constrained open systems by forming new or adapting existing collaboration communities.	
\end{conf-talk}

\begin{conf-talk}[]
	{Model-driven Self-optimization for Energy-efficient Software}
	{Sebastian Götz}
	{University of Technology Dresden, DE}
	\indexauthors{Götz!Sebastian}
	\noindent
In my talk I will give an overview of my recent work on model-driven software development at runtime, with a particular focus on self-optimization aiming at energy-efficiency and its application in robotics.	
\end{conf-talk}

\begin{conf-talk}[]
	{Towards Self-Aware and Self-Adapting Performance Models}
	{Johannes Grohmann}
	{University of Würzburg, D}
	\indexauthors{Grohmann!Johannes}
	\noindent
Performance models are possible components of self-aware computing systems, as they allow such systems to reason about their own state and behavior. Our research is targeted towards such performance models. We propose an approach to meta-self-awareness, making the processes of model creation, maintenance and solution themselves self-aware. This enables the automated selection and adaption of software performance engineering approaches specifically tailored to the system under study.	
\end{conf-talk}

\begin{conf-talk}[]
	{Exploiting Model-driven Engineering in Robotics at Design Time and Run Time}
	{Nico Hochgeschwender}
	{Université du Luxembourg, LU}
	\indexauthors{Hochgeschwender!Nico}
	\noindent
Engineering advanced autonomous systems such as robotic applications is a knowledge-intensive process that reflects, involves and builds upon decisions from complex, heterogenous fields of research and engineering -- reaching from hardware design, domains such as control, perception or planning to software engineering. Although the latest advancements in those fields contributed significantly to the development of sophisticated applications, robots' task spectrum and autonomy often remains limited to carefully engineered applications. One of the reasons is that in robotics software engineering in particular, the integration of those fields is all too often solved in an ad-hoc manner, where knowledge and assumptions about the robot's software remains implicit. Model-driven engineering aims to remedy this situation by introducing modeling languages to capture this knowledge explicitly and formally in the form of various domain models. However, these models are merely seen as a way to support humans during the robot's software design process. In this talk I will argue and demonstrate that robots themselves should be first-class consumers of this knowledge. Having this knowledge enables robots to autonomously adapt their software to the various and changing run-time requirements induced, for example, by the robot's task or environment.
\end{conf-talk}

\begin{conf-talk}[]
	{Making the Everyday Life smarter through Cyber-physical Systems}
	{Christian Krupitzer}
	{University of Mannheim, DE}
	\indexauthors{Krupitzer!Christian}
	\noindent
Cyber-physical systems seamlessly intertwine physical everyday objects with virtual software to provide intelligent, adaptive, and connected services to users. These systems enable new types of applications. This presentation gives an overview on several research projects in the cyber-physical systems domain. The iCOD project presents an approach for platooning coordination, i.e. coordination of self-driving vehicles to convoys on highways. In the adaptive authentication, we envision a password-free world through a distributed, adaptive system for authentication. The third project investigates the application of a layered meta-model for engineering adaptive systems, predictive maintenance, and adaptive communication protocols in the domain of Industry 4.0. Last, in cooperation with the soccer club TSG Hoffenheim, we research the use of data mining and virtual reality in sports.	
\end{conf-talk}

\begin{conf-talk}[]
	{Providing Resilience and Efficiency to the Internet of Things}
	{Ingrid Nunes}
	{Universidade Federal do Rio Grande do Sul, BR}
	\indexauthors{Nunes!Ingrid}
	\noindent
Since the popularisation of the Internet, software systems have evolved from simple standalone applications running in isolated computers to distributed systems that largely interact with each other running in various types of devices. The network that emerges from all these communicating devices is referred to as Internet of Things (IoT). Because of the distribution of software components and the dependency among them, which are not necessarily part of the same organisation, ensuring system resilience and efficiency becomes a challenge. In this talk, I'll introduce work that my research group has been developing towards the use of the autonomous components to provide these quality attributes to such systems. Resilience is achieved by means of a set of techniques that allow system components, implemented as intelligent agents, to react with remediation actions to situations that may compromise the functioning of the system to later seek and autonomously resolve causes of problems that occur at runtime. Efficiency, in turn, is improved by an adaptive application-level framework, able to autonomously and manage cache data at a method-level. Recent publications on these two topics can be seen at \url{http://inf.ufrgs.br/~ingridnunes}.	
\end{conf-talk}

\begin{conf-talk}[]
	{Control of things}
	{Alessandro Vittorio Papadopoulos}
	{MDH, Västerås, SE}
	\indexauthors{Papadopoulos!Alessandro V.}
	\noindent
Nowadays, we live in a society with billions of devices that are interconnected, and interact together to improve the quality of our lives. The management and processing of information and knowledge have by now become our main resources, and the fundamental factors of economic and social development. This has been possible by the recent advances in computing systems, ranging from embedded devices to cloud computing systems, but it poses a number of challenges in the management of the emerging complexity. In order to tame such a complexity, mathematically grounded approaches are needed, in order to design next generation computing systems. This talk discusses how control theory can be used for designing efficient solutions for controlling the behavior of computing systems, and presents successful applications, together with current challenges and opportunities in the field.	
\end{conf-talk}

\begin{conf-talk}[]
	{Approaching a Self-Adaptive Middleware for Network Adaptations}
	{Martin Pfannemüller}
	{University of Mannheim, DEil}
	\indexauthors{Pfannemüller!Martin}
	\noindent
Self-adaptive capabilities reduce maintenance effort and help to reconfigure systems at runtime according to changes in their context. With the increasing number of computation devices induced by trends such as smart home and Industry 4.0, managing and adapting the network gets gradually more important. The goal of this work is to approach a self-adaptive middleware for network adaptations which is able to change the network behavior at runtime. The middleware should provide an abstraction for specifying network adaptations on a higher level with support for different knowledge sources as well as diverse target systems such as software-defined networking.	
\end{conf-talk}

\begin{conf-talk}[]
	{Engineering the aggregate}
	{Danilo Pianini}
	{University of Bologna, IT}
	\indexauthors{Pianini!Danilo}
	\noindent
A distributed system can be seen as a single computational machine rather than a collection of multiple communicating machines, as it is usually perceived. Reasoning on the aggregate of situated devices under this privileged point of view can lead to interesting engineering solutions that allow for abstracting away the networking protocols, and focusing on producing advanced, self-stabilizing coordination algorithms.	
\end{conf-talk}

\begin{conf-talk}[]
	{The New Abstraction: Engineering Search Spaces for Machine Learning}
	{Barry Porter}
	{Lancaster University, UK}
	\indexauthors{Porter!Barry}
	\noindent
Humans have long struggled with the increasing complexity and scale of software. Baseline complexity, measured in lines of code, is compounded by the challenges of highly dynamic deployment environments, where behavioural adaptation is needed to offer good service. In this talk we argue that raising abstraction levels in traditional software engineering approaches has reached its limit in addressing complexity, and a new methodology is required: an abstraction of designing search spaces for machine learning. We examine this idea through two major areas: the component model that powers emergent software systems and how it offers a universal substrate for runtime behavioural search; and the challenge of automated program logic synthesis to populate a component library without the human programmer. In both areas, the engineering problem is fundamentally shifted from designing abstraction layers to instead designing a search space for machine learning; we explore the properties of this process and how we might generalise it into common practice throughout autonomous systems engineering.	
\end{conf-talk}

\begin{conf-talk}[]
	{Self-adaptive Learning in Decentralized Combinatorial Optimization}
	{Evangelos Pournaras}
	{ETH Zurich, CH}
	\indexauthors{Pournaras!Evangelos}
	\noindent
The democratization of Internet of Things and ubiquitous computing equips citizens with phenomenal new ways for online participation and decision-making in application domains of smart grids and smart cities. When agents autonomously self-determine the options from which they make choices, while these choices collectively have an overall system-wide impact, an optimal decision-making turns into a combinatorial optimization problem known to be NP-hard. This paper contributes a new generic self-adaptive learning algorithm for a fully decentralized combinatorial optimization: I-EPOS, the Iterative Economic Planning and Optimized Selections. In contrast to related algorithms that simply parallelize computations or big data and deep learning systems that often require personal data and overtake of control with implication on privacy-preservation and autonomy, I-EPOS relies on coordinated local decision-making via structured interactions over tree topologies that involve the exchange of entirely local and aggregated information. Strikingly, the cost-effectiveness of I-EPOS in regards to performance vs. computational and communication cost highly outperforms other related algorithms that involve non-local brute-force operations or exchange of full information. The algorithm is also evaluated using real-world data from two state-of-the-art pilot projects of participatory sharing economies: (i) energy management and (ii) bicycle sharing. The contribution of an I-EPOS open source software suite implemented as a paradigmatic artifact for community aspires to settle a knowledge exchange for the design of new algorithms and application scenarios of sharing economies towards highly participatory and sustainable digital societies.	
\end{conf-talk}

\begin{conf-talk}[]
	{From Self-adaptation to Self-composition: Transcending Autonomic Computing Limitations}
	{Roberto Rodrigues Filho}
	{Lancaster University, UK}
	\indexauthors{Filho!Roberto Rodrigues}
	\noindent
Contemporary systems are increasingly complex. The reason vary from their size, heterogeneous infrastructures and highly volatile operating environments. As a response to this complexity, the area of Autonomic Computing (AC) has gained significant importance. However, the focus of AC research has long been on engineered self-adaptation, in which human engineers determine how and where to adapt a system’s structure and/or parameters to accommodate changes. We argue for a shift of focus from engineered adaptation to autonomous composition as a way to transcend limitations of current approaches, and to build real-world autonomous everyday software. In this talk, we discuss how to autonomously compose a web platform, and the decisions involved in the composition process. To realise local systems this involves autonomous decisions of which small components should be composed, and how, at each moment to deliver the desired system; to realise distributed systems this involves autonomous decisions on which local components should be relocated and/or replicated amongst machines. We end the talk inviting the community to join us to further explore this paradigm shift, and to push the concepts of AC to effectively realise multi-purpose, everyday software systems.	
\end{conf-talk}

\begin{conf-talk}[]
	{ECo-IoT: an Architectural Approach for Realizing Emergent Configurations in the Internet of Things}
	{Romina Spalazzese}
	{Malmö University, SE}
	\indexauthors{Spalazzese!Romina}
	\noindent
The rapid proliferation of the Internet of Things (IoT) is changing the way we live our everyday life and the society in general. New devices get connected to the Internet every day and, similarly, new IoT services and applications exploiting them are developed across a wide range of domains. The IoT environment typically is very dynamic, devices might suddenly become unavailable and new ones might appear. Similarly, users enter and/or leave the IoT environment while being interested in fulfilling their individual needs. These key aspects must be considered while designing and realizing IoT systems. In this talk I will discuss ECo-IoT, an architectural approach to enable the automated formation and adaptation of Emergent Configurations (ECs) in the IoT. An EC is formed by a set of things, with their services, functionalities, and applications, to realize a user goal. ECs are adapted in response to (un)foreseen context changes e.g., changes in available things or due to changing or evolving user goals. In the talk, I will discuss: (i) an architecture and a process for realizing ECs; and (ii) a prototype we implemented for (iii) the validation of ECo-IoT through an IoT scenario.	
\end{conf-talk}

\begin{conf-talk}[]
	{Increased system autonomy through learning and self-* properties}
	{Sven Tomforde}
	{University of Kassel, DE}
	\indexauthors{Tomforde!Sven}
	\noindent
The goal of my research activities is to increase the autonomy of technical systems by learning and self-* properties. The basic theme is to master large-scale interconnected systems through mechanisms such as self-adaptation and self-organisation - which requires autonomous learning capabilities. In order to achieve this goal, four research areas are distinguished: ``autonomous machine learning'', ``data-driven system modelling'', ``distributed control algorithms'' and ``computational trust and security mechanisms'', resulting in different methods, tools and applications for distributedly interacting, intelligent and autonomous systems. As application areas, I focus on traffic control, industry automation, smart energy grid, and data communication networks.	
\end{conf-talk}

\begin{conf-talk}[]
	{Self-Adaptive Search for Sapienz}
	{Thomas Vogel}
	{HU Berlin, DE}
	\indexauthors{Vogel!Thomas}
	\noindent
In this talk, I will outline how feedback and self-adaptation can improve search heuristics in search-based software engineering. The focus will be on using results of fitness landscape analysis to dynamically adapt the search when generating test suites for Android apps with Sapienz, a search-based testing tool for apps.	
\end{conf-talk}

\chapter[Participants]{Participants}

\begin{table}[htbp]
	\centering
	{\footnotesize
	\begin{tabular}{p{55mm}p{65mm}p{55mm}}
		= Aimee Borda \newline Trinity College Dublin, IR &
		= Fatemeh Golpayegani\newline Trinity College Dublin, IR&
		= Martin Pfannemüller  \newline University of Mannheim, DE\\ \ \\

		= Christian Cabrera  \newline Trinity College Dublin, IR&
		= Sona Ghahremani \newline Hasso Plattner Institute, DE&
		= Danilo Pianini \newline University of Bologna, IT\\ \ \\
				
		= Mirko D'Angelo\newline Linnaeus University, SE&
		= Sebastian Götz \newline University of Technology Dresden, DE&
		= Barry Porter \newline Lancaster University, UK\\ \ \\

		= Ada Diaconescu \newline Télécom ParisTech, FR&
		= Johannes Grohmann  \newline University of Würzburg, DE&
		= Evangelos Pournaras  \newline ETH Zurich, CH\\ \ \\

		= Lukas Esterle \newline Aston University, UK&
		= Nico Hochgeschwender \newline Université du Luxembourg, LU&
		= Roberto Rodrigues Filho \newline Lancaster University, UK\\ \ \\
		
		= Erik Fredericks \newline Oakland University, USA&
		= Christian Krupitzer  \newline University of Mannheim, DE&
		= Romina Spalazzese \newline Malmö University, SE\\ \ \\		

		= Sylvain Frey\newline DeepMind, UK&
		= Ingrid Nunes \newline Univ. Federal do Rio Grande do Sul, BR&
		= Sven Tomforde \newline University of Kassel, DE\\ \ \\

		= Simos Gerasimou \newline University of York, UK&
		= Alessandro V. Papadopoulos \newline MDH, Västerås, SE&
		= Thomas Vogel \newline HU Berlin, DE\\ \ \\

		= Ilias Gerostathopoulos \newline Technical University Munich, DE&&
		\\
		
	\end{tabular}
}\end{table}

\begin{center}
	\includegraphics[width=.74\linewidth]{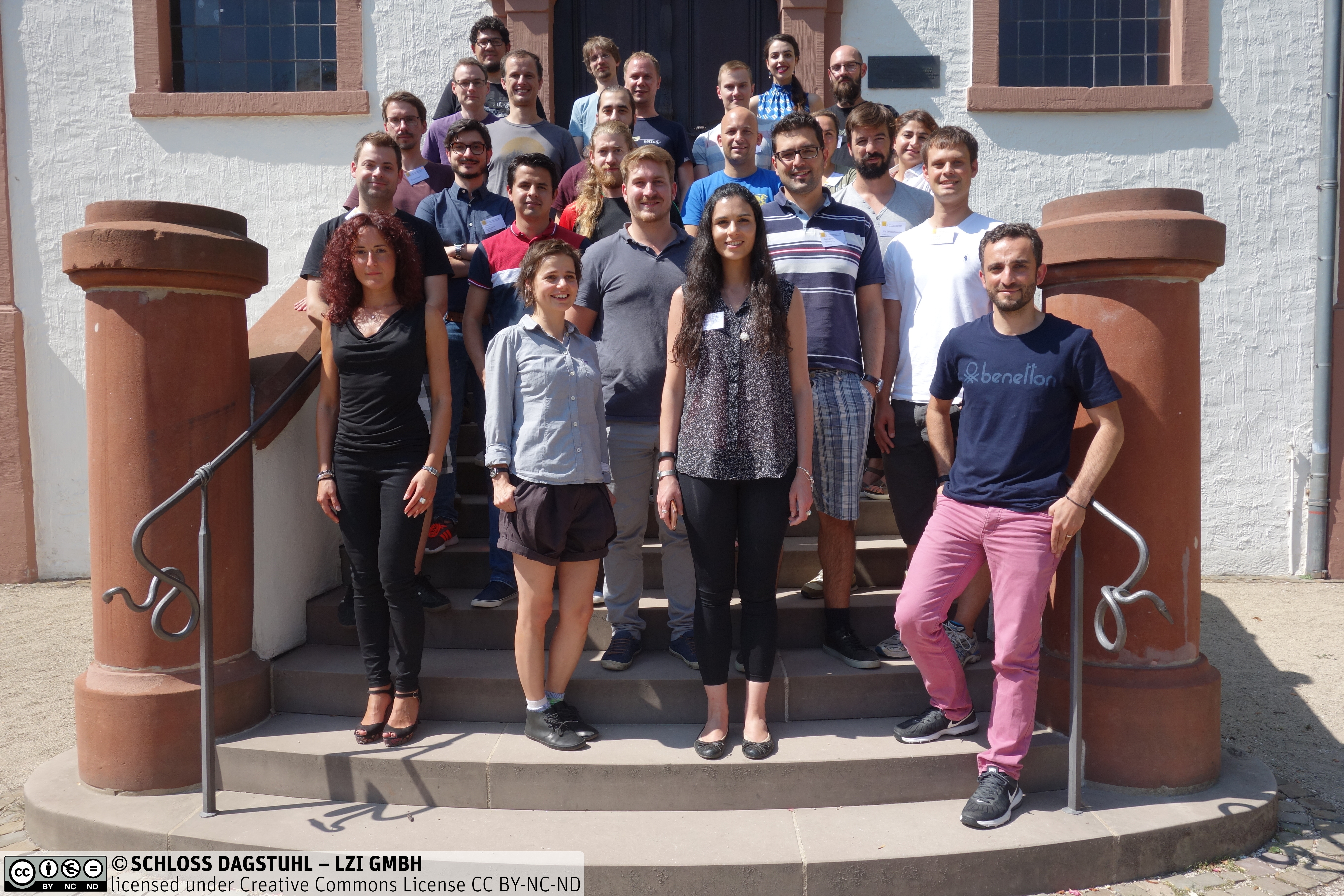}
\end{center}

\backmatter
\renewcommand{\indexname}{Author Index}
\printindex

\end{document}